# Generative AI for Software Project Management: Insights from a Review of Software Practitioner Literature

*Lakshana Iruni Assalaarachchi, Monash University (https://orcid.org/0009-0006-1848-3233)*

*Zainab Masood, Prince Sultan University (https://orcid.org/0000-0002-3592-8250)*

*Rashina Hoda, Monash University (https://orcid.org/0000-0001-5147-8096)*

*John Grundy, Monash University (https://orcid.org/0000-0003-4928-7076)*

## Abstract

Software practitioners are discussing GenAI transformations in software project management openly and widely. To understand the state of affairs, we performed a grey literature review using 47 publicly available practitioner sources including blogs, articles, and industry reports. We found that software project managers primarily perceive GenAI as an "assistant", "copilot", or "friend" rather than as a "PM replacement", with support of GenAI in automating routine tasks, predictive analytics, communication and collaboration, and in agile practices leading to project success. Practitioners emphasize responsible GenAI usage given concerns such as hallucinations, ethics and privacy, and lack of emotional intelligence and human judgment. We present upskilling requirements for software project managers in the GenAI era mapped to the Project Management Institute's talent triangle. We share key recommendations for both practitioners and researchers.

## Introduction

Generative AI (GenAI) is starting to show the potential to transform Software Engineering (SE), facilitating successful project completions [1]. This transformation requires reshaping the roles and skills of software practitioners to facilitate better Human-AI collaboration. Software Project Manager (PM) is one such role which is evolving with the advent of GenAI. Repetitive tasks of software PMs, such as creating and maintaining project documents, optimizing resource allocation, and predicting risks associated with projects, may now be automated using GenAI [2], [3]. Some studies propose AI-based chatbots or prompting experiments using tools like ChatGPT for software project management (SPM) activities [4], [5], [6].

The 2023 Project Management Institute (PMI) report revealed that 21% of PM professionals are using AI for project management, while 58% of respondents said AI will have a "major" or "transformative" impact in the future [7]. The 2024 PMI survey, particularly focusing on Generative AI in Project Management, indicates that 20% of respondents use GenAI in more than half of projects ("trailblazers") while 54% use in 16-50% of projects as it assists successful project delivery [8]. A systematic literature review of academic publications between 2017 to 2024 on Large Language Models (LLMs) for SE identified application of LLMs in SPM as a less studied phenomenon [9]. While research into the topic has barely begun, practitioners are freely discussing their experiences of GenAI usage in SPM on social platforms such as LinkedIn, blogs, and through contributing to industry reports. There is much real-world evidence and knowledge to be gained from these sources on how software practitioners are using GenAI[1] in SPM activities, what concerns they face when using GenAI in SPM, and whether and how software PMs need to upskill themselves in the GenAI era. Therefore, we conducted a review of practitioner literature sources (also known as a grey literature review) to answer the following research questions:

- ***RQ1:*** *How do software practitioners use GenAI in SPM and what benefits do they report?*
- ***RQ2:*** *What concerns do software practitioners have when using GenAI in SPM?*
- ***RQ3:*** *How are software PMs upskilling themselves in the GenAI era?*

Through our review, we hope to present timely insights and recommendations on the beneficial use cases of GenAI in SPM, practitioners' concerns when using it, along with strategies to overcome them, and skills required by software PMs in the GenAI era. Software practitioners can use these insights and recommendations to keep themselves up to date in this GenAI era, while gaining a competitive advantage in their career as GenAI becomes pervasive in modern workplaces.

## Methodology

We carried out a grey literature review using blog posts, articles on professional networks such as LinkedIn, and industry reports that are publicly available and freely accessible on the internet. Google was used as the search tool, as it has been adopted by similar previous studies [10]. We got 618,000 results after expanding the search string (Figure 1), with

---

[1] For this study we considered GenAI to include prompt-based tools such as ChatGPT used for software project management (SPM) artifacts (e.g., texts, images, codes, templates, etc.) and also GenAI integrations to SPM tools such as Jira which provide recommendations.

keywords such as "Large Language Models", "LLMs", "ChatGPT", and "Co-Pilot" which are some of the most popular GenAI tools currently [10].

We later filtered the initial results to 38,700 applying our inclusion criteria: being published between December 2022 (after ChatGPT public release) to May 2025 (current study date) and the language being English. We narrowed them further to 15,400 results by filtering only the LinkedIn articles, blogs, and reports that have sufficient details for thematic analysis while ignoring short posts or comments. We manually reviewed only the first five pages of Google results as they reflect the most relevant results (with rapidly decreasing relevance on additional pages). We considered only the practitioner literature sources written by software practitioners, software firms, or PM related organizations (identified through affiliation or professional qualifications mentioned in the source or through LinkedIn profile) to ensure context relevancy and avoid promotion of specific tools. The selected practitioner sources were then subjected to qualitative thematic analysis to derive key themes from the practitioners' experiences and perceptions on GenAI usage in SPM (Figure 1). Our detailed study procedure with inclusion and exclusion criteria, the codebook, and grey literature sources[2] are available in our *supplementary file*[3].

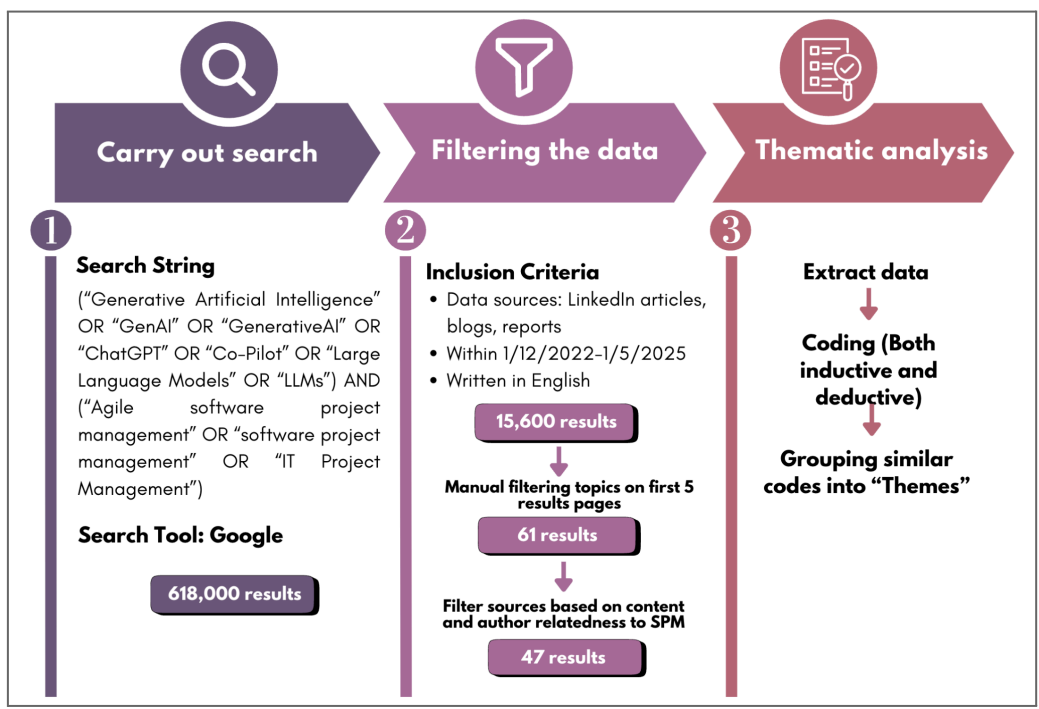

**Figure 1: Overview of grey literature review methodology**

---

[2] Ethics approval for this study was secured from the Monash University Human Ethics Review Committee. Findings are presented in an aggregated format as required in our ethical approval.
[3] Supplementary file: https://zenodo.org/records/17282415

# Key Findings
## Demographic Data

Figure 2(A) indicates the number of practitioner sources across major discussion topics and a summary of findings particular to those topics[4]. Figure 2(B) depicts the number of practitioner sources over the years considered for our review period. Most sources discuss GenAI usage in SPM through LinkedIn articles (Figure 2(C)). A majority discuss the impact of GenAI on SPM based on their own or their organization's experience, while some refer to learnings from industry reports and professional courses (Figure 2(D)).

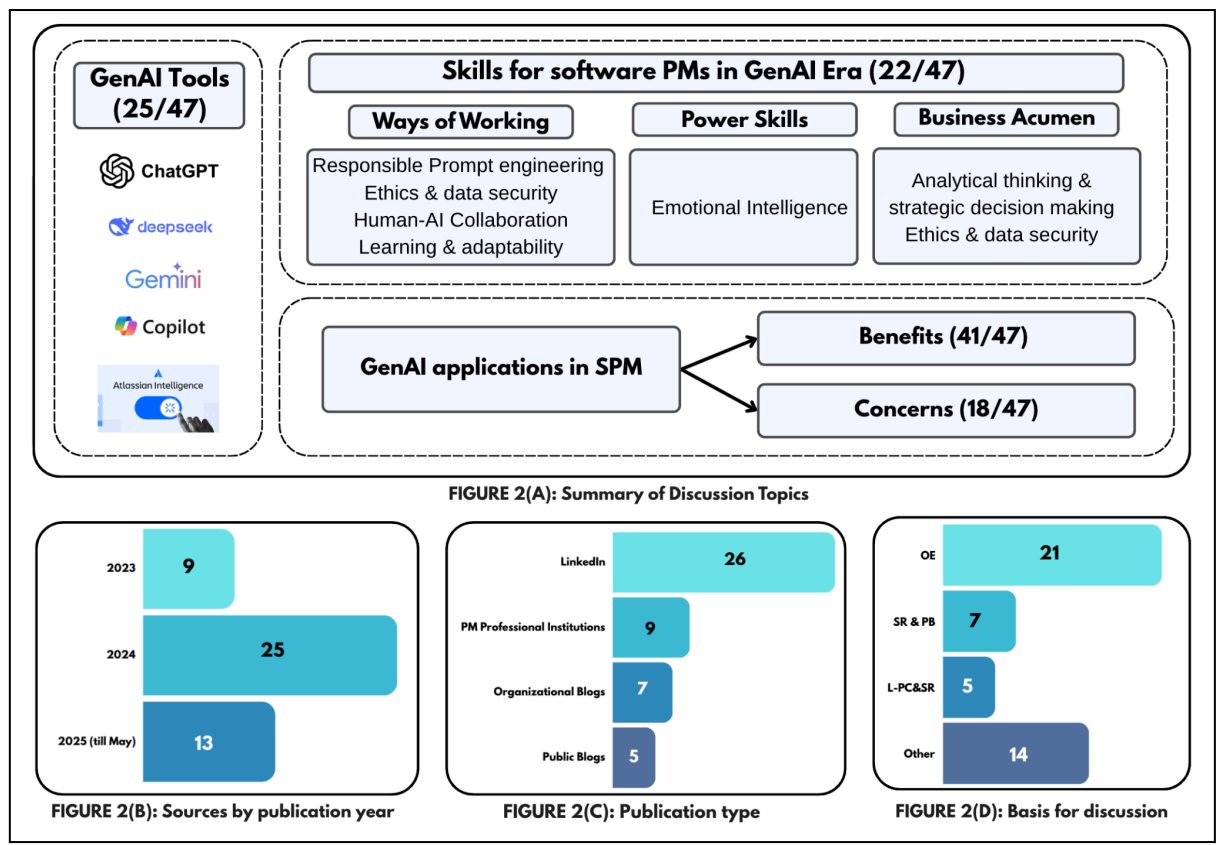

**Figure 2: Distributions - Sources by publication year, publication type, basis for discussion, and summary of discussion topics.**
*Note: In Figure 2(D), OE = Own Experience, SR & PB = Survey Reports & Professional Blogs, L-PC&SR = Learnings from Professional Courses & Survey Reports, Other = Sources with unclear basis*

## RQ1: How do software practitioners use GenAI in SPM and what benefits do they report?

We identified that 25 out of 47 sources mention the use of GenAI tools such as ChatGPT, Microsoft Copilot, and Gemini in SPM activities. Meanwhile, SPM tools such as Jira, MS Project, Asana, Trello, and Slack were also reported to feature GenAI integrations (e.g., Atlassian Intelligence, Rovo and Slack AI). Practitioners expressed their willingness to use

---
[4]Note that some sources cover more than one topic in that article, blog or report.

GenAI within the familiar user interfaces of their SPM tools rather than shifting to standalone GenAI tools.

We noted that 17 practitioner sources view GenAI as "an assistant", "a co-pilot", "a team member", or "a friend" supporting them in their day-to-day activities. The benefits reported included time savings, improved efficiency, faster deliverables through automation of routine tasks (e.g., generating SPM documents such as project plans, project reports, meeting minutes, and follow-up emails), better decision making with predictive analytics, improved communication and collaboration, and project success through effective risk, cost, and time management.

We also noted GenAI usage to support agile practices, partly fulfilling the idea of augmented agile where agile practices were envisioned to be augmented with AI while incorporating human-centered values [11]. We know that agile practices such as story point estimation and tasks allocations for sprints are meant to be human-centered and collaborative, with due consideration of team dynamics, skills, and the workloads of each team member [4], [11], [12]. However, currently there is little guidance on how GenAI use can be more human-centered and collaborative. Therefore, PMs should carefully assess GenAI applications' suitability for such collaborative tasks and its impact on team dynamics.

One practitioner source described their experience in using GenAI as a central virtual assistant in most SPM tasks. This suggests the use of GenAI as a lightweight SPM assistant, although it cannot fully replace enterprise SPM tools in terms of data storage, sharing, and collaboration.

Overall, combining usage trends found in this review with recommendations from established practitioner and research literature, we recommend that practitioners should assess the suitability of GenAI tools for collaborative tasks, and use these tools responsibly, avoiding over-reliance [1], [4].

**RQ2: What concerns do software practitioners have when using GenAI in SPM?**

In addition to the reported benefits, we found that 18 of the 47 practitioner sources discussed potential concerns from the use of GenAI in SPM. Practitioners must be aware of and address these concerns carefully to avoid possible negative consequences, as detailed below.

- **Ethical and privacy concerns:** 14 practitioner sources reported that software practitioners encounter privacy concerns as they often share project-specific,

      client-specific, or organizational information when providing the context to GenAI tools. To overcome this concern, we recommend that practitioners should become aware of the relevant national and organizational guidelines and client policies on data sharing before sharing data with GenAI. Where required, share data iin an anonymised or de-identified form when using GenAI for project works to avoid ethical concerns [4].

- **Data quality and accuracy:** Nine practitioner sources elaborated on experiencing hallucinations in GenAI outcomes which were seen to cause inaccurate insights and decisions. They recommend practitioners to always review the GenAI outcomes to avoid such consequences. Meanwhile, five practitioner sources mention poor prompts as a cause of low-quality outcomes. This highlights the need to master prompt engineering skills and being the human-in-the-loop to review GenAI outcomes, similar to recent research literature [1], [4].

- **Lack of Emotional Intelligence and Human Judgment:** Inability of GenAI to have empathy, emotion, and human intuition in their outcomes similar to humans was reported in four sources. Therefore, we recommend practitioners not to use GenAI as a replacement for human judgment, especially in scenarios where more human involvement is required, such as conflict resolution and mentoring team members.

Interestingly, none of the practitioner sources raised energy consumption and sustainability concerns. A recent Google survey identified that 0.24 watt-hours of electricity (equivalent to operating a microwave for one second) is consumed by a typical Gemini query [13]. Therefore, we recommend that practitioners should use GenAI responsibly, without simply using it for every task, to reduce energy consumption which is a critical sustainability concern.

Table 1 summarizes the GenAI applications mapped to their proposed and potential benefits mentioned by 41 practitioner sources (RQ1) and associated concerns (RQ2) mentioned by 18 practitioner sources.

**Table 1: GenAI applications in Software Project Management (SPM), their benefits (n=41/47), and concerns (n=18/47)**

| GenAI application in SPM | Description | Benefits | Concerns |
|---|---|---|---|
| Automation of Routine Tasks (n=38) | Use of GenAI to automate less complex and routine tasks of software PMs. *(e.g., generate various documents such as project plans, project reports, meeting minutes, and follow-up emails)* | Save time for strategic tasks *(e.g., decision making, managing stakeholders)* Increase efficiency and productivity Faster Deliverables | Privacy concerns with sharing sensitive project information to generate SPM documents Data quality and accuracy in generated SPM documents Lack of emotions and human judgment in generated outcomes (e.g., tone of email) |
| Predictive Analytics (n=30) | Use of GenAI to support analysis of historical and current data to identify patterns and provide future insights during risk prediction, cost estimations, schedule forecasts and resource allocations. | Data-driven decision making Better resource allocation Project success through effective risk, cost and time management | Privacy concerns with sharing sensitive project information for analysis Data quality and accuracy in analyzed data Lack of emotional intelligence and human judgment (e.g., not considering team dynamics, experiences) |
| Communication and Collaboration (n=22) | Use of GenAI and GenAI-powered Chatbots to answer frequently asked questions on project status and GenAI integrations to communication tools such as Slack and SPM tools such as Jira | Enhance communication Better task tracking Ease collaboration Better workflow management | Privacy concerns with sharing sensitive project information Data quality and accuracy (e.g., inaccurate information in FAQ answer) Lack of emotions and human judgment in outcomes (e.g., tone of email) |

| Assist in agile PM practices (n=10) | Use of GenAI assistance in agile practices such as backlog grooming, create clear and accurate user stories in shorter time, story point estimations, sprint planning and retrospectives. | Enhance agile project success | Privacy concerns with sharing sensitive project information in agile practices Data quality and accuracy (e.g., inaccurate story point estimations) Lack of emotional intelligence and human judgment in outcomes (e.g., not considering team dynamics in story point estimations) |
|---|---|---|---|

**RQ3: How are software PMs upskilling themselves in the GenAI era?**

We identified 22 practitioner sources discussing the skills that need to be improved in the GenAI era to get a competitive advantage in their careers, along with upskilling sources. We have mapped these skills to the latest PMI Talent Triangle [14], a professional skills framework (Figure 3) that highlights three categories:

- **Ways of working**, which refers to various methods, practices, and tools used in SPM workflows to deliver effective outcomes. Practitioner reported skills such as *prompt engineering, ethics and data security, human-AI collaborations,* and *learning and adaptability* can be mapped to ways of working because these skills are necessary to learn and adopt GenAI effectively and responsibly in SPM workflows to deliver successful projects.
- **Power skills**, which refers to interpersonal skills that supports better stakeholder management. *Emotional intelligence* is mapped as a power skill, since it can strengthen the PM's ability to maintain better relationships with stakeholders, while GenAI cannot apply emotional intelligence in ways human do.
- **Business acumen**, which refers to having project-specific domain knowledge and effective decision making aligned with organizational strategic goals. *Analytical thinking and strategic decision making,* and *ethics and data security knowledge* about project-specific domain is mapped as business acumen since they support effective decision making aligned to strategic business goals, and compliance to ethical and legal GenAI obligations in project specific domain.

*Upskilling sources for PMs in GenAI era*

Practitioner sources propose online blogs, videos, documentations and certifications courses as main upskilling sources for software PMs in GenAI era (e.g., Section 4, Page 9 in *supplementary file*). Practitioner sources also mention that the GenAI tool can itself become a good learning platform for software PMs. They can ask questions from the GenAI tool on areas for learning, such as *"how to write better prompts in generating a project plan?", "how to master emotional intelligence?", or "resources to improve analytical thinking in software project management"*, along with hands-on experience.

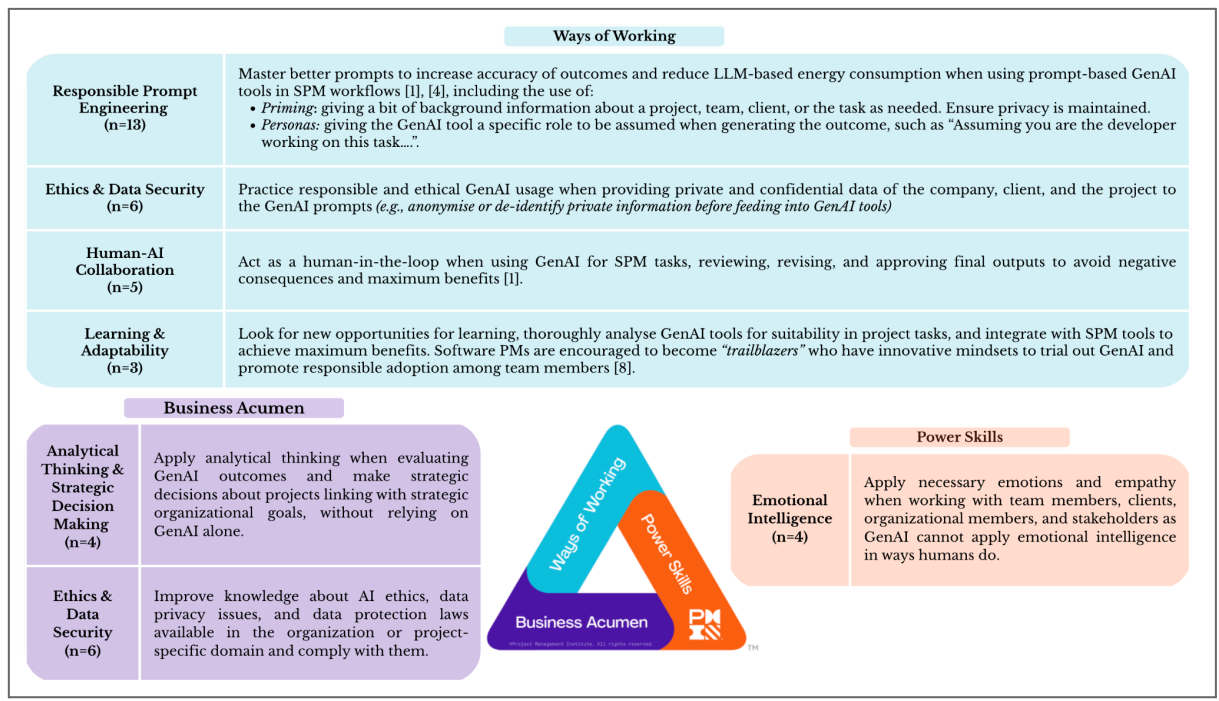

**Figure 3: Upskilling requirements for Software PMs in GenAI Era (n=22/47) mapped to PMI Talent Triangle [14]**

## Key Recommendations

**For Practitioners**

- **Upskill for responsible GenAI usage:** Master prompt engineering to get accurate outcome, and enhance knowledge on ethics and data security for responsible GenAI usage by following professional and other learning sources (e.g., Section 4, Page 9 in *supplementary file*).
- **Practice and promote responsible use of GenAI:**
  - Use GenAI as an assistant in less-complex routine tasks after assessing the suitability for the task.

- - Always act as the human-in-the-loop and review GenAI outcomes thoroughly, especially in strategic tasks that need more human involvement (e.g., strategic planning, and conflict resolution), to address issues with hallucinations.
  - Adhere to national, organizational, project-specific, and client's GenAI usage policies, when using GenAI platforms, including data sharing.
  - Provide data in an anonymised or de-identified form when sharing sensitive information.
  - Acknowledge use of GenAI in SPM documents or decisions to ensure transparency.
- **Select appropriate tools:** Experiment with different GenAI tools for the same use case (e.g., task size/effort estimation) to identify the most suitable GenAI tool for the particular use case. Encourage team member feedback to benefit from multiple perspectives.
- **Maintain a prompt library:** Create a prompt library shared with team members (e.g., Google Sheet) to be used in repetitive tasks and regularly update with team inputs.
- **Organizations are encouraged to develop responsible GenAI usage guidelines and conduct training** on responsible use of GenAI for staff to assist in successful project delivery.

**For Researchers**

- **Researchers are encouraged to conduct empirical research on:**
  - Impact of GenAI adoption on software (agile) project management practices and project success
  - Barriers to and challenges of GenAI adoption in SPM along with strategies to overcome them
  - Ethical and responsible use of GenAI in SPM, including role of human values and (un)desirable consequences
  - The evolving role of software PM in the GenAI era
  - Understanding human expectations, experiences, and impact of Human-AI collaboration in SPM

## Conclusion

**Limitations of our study**

Our study is limited only to text-based and publicly available articles, blog posts, and reports about the GenAI for SPM. While these sources provided valuable insights, we acknowledge the limitations in the exclusion of audio and video data sources, given their scarcity and data quality issues. As GenAI is still evolving in SPM, we could consider only 47 valid data sources that might increase over the years with different opinions of other practitioners, and future research might reveal wider findings on the use of GenAI for SPM. Despite these limitations, we were able to present insights for practitioners and researchers on the beneficial GenAI applications for SPM, possible concerns and strategies to overcome them, and upskilling requirements with sources. We believe these findings will build the necessary foundation for them to explore GenAI for SPM further.

**Toward Agentic SPM**

Beyond GenAI, with advancement in AI models and technologies, agentic SPM is expected to impact SPM where AI agents are expected to proactively augment human decision making through real-time data monitoring, better risk management, and autonomous task management. Rather than a replacement, SPM agents are envisioned as a companion for the software PM, saving time for strategic activities, providing ethical oversight, boosting stakeholder engagements, and enchaining efficiency in SPM workflows [15]. How well these predictions come to pass and what desirable and undesirable consequences they may bring remains to be seen. Overall, we envision a new SPM era where software PMs are not limited to management of people and processes, where the lines between coding and managing are blurred, and where they are transformed into navigators of responsible human-AI collaboration.

## References

[1]     T. Şimşek, Ç. Gülşeni, and G. A. Olcay, "The Future of Software Development with GenAI: Evolving Roles of Software Personas," IEEE Engineering Management Review, pp. 1–8, Sep. 2024, doi: 10.1109/EMR.2024.3454112.

[2]     M. Kanbur, O. P. C, and P. Kulkarni, "Creative AI in Software Project Management", in 2023 2nd International Conference on Futuristic Technologies (INCOFT), IEEE, 2023, pp. 1–9. doi: 10.1109/INCOFT60753.2023.10425234.

[3]     L. Song and L. L. Minku, "Artificial Intelligence in Software Project Management", in Optimising the Software Development Process with Artificial Intelligence, Springer Nature Singapore, 2023.

[4]     B. Haidabrus, "Generative AI in Agile, Project, and Delivery Management", in Proceedings of the 7th International Conference on Design, Simulation, Manufacturing: The Innovation Exchange, DSMIE-2024, Springer Nature Switzerland, 2024. doi: 10.1007/978-3-031-61797-3_9.

[5]     K. Alam, M. H. Bhuiyan, M. S. Islam, A. H. Chowdhury, Z. A. Bhuiyan, and S. Ahmmed, "Co-Pilot for Project Managers: Developing a PDF-Driven AI Chatbot for Facilitating Project Management", IEEE Access, vol. 13, pp. 43079–43096, 2025, doi: 10.1109/ACCESS.2025.3548519.

[6]     A. Bahi, J. Gharib, and Y. Gahi, "Integrating generative AI for advancing agile software development and mitigating project management challenges," International Journal of Advanced Computer Science and Applications, vol. 15, no. 3, Jan. 2024, doi: 10.14569/ijacsa.2024.0150306.

[7]     Project Management Institute (PMI), "Shaping the Future of Project Management With AI". Accessed: May 25, 2025. [Online]. Available: https://www.pmi.org/learning/thought-leadership/ai-impact/shaping-the-future-of-project-management-with-ai

[8]     Project Management Institute (PMI), "First Movers' Advantage: The Immediate Benefits of Adopting Generative AI For Project Management". Accessed: May 25, 2025. [Online].                                                                                                                                             Available: https://www.pmi.org/learning/thought-leadership/benefits-of-ai-for-project-management

[9]     X. Hou et al., "Large Language Models for Software Engineering: A Systematic Literature Review", ACM Transactions on Software Engineering and Methodology, vol. 33, no. 8, pp. 1–79, 2024, doi: 10.1145/3695988.

[10]    N. Davila, J. Melegati, and I. Wiese, "Tales From the Trenches: Expectations and Challenges From Practice for Code Review in the Generative AI Era", IEEE Softw, vol. 41, no. 6, pp. 38–45, 2024, doi: 10.1109/MS.2024.3428439.

[11]    R. Hoda, H. Dam, C. Tantithamthavorn, P. Thongtanunam, & M. A. Storey, "Augmented agile: Human-centered AI-assisted software management," IEEE Software, 40(4), pp. 106-109, 2023, doi: 10.1109/MS.2023.3268725.

[12]    P. Diebold, "From backlogs to bots: Generative AI's impact on agile role Evolution," Journal of Software Evolution and Process, Nov. 2024, doi: 10.1002/smr.2740.

[13]    C. Crownhart, "In a first, Google has released data on how much energy an AI prompt uses," MIT Technology Review, Aug. 21, 2025. [Online]. Available: https://www.technologyreview.com/2025/08/21/1122288/google-gemini-ai-energy/.

[14]    PMI, "Plan Your Development to the PMI Talent Triangle | PMI," Pmi.org, 2024. https://www.pmi.org/certifications/certification-resources/maintain/talent-triangle.

[15]    L. Hughes, Y. K. Dwivedi, K. Li, M. Appanderanda, M. A. Al-Bashrawi, and I. Chae, "AI Agents and Agentic Systems: Redefining Global IT Management," Journal of Global Information Technology Management, vol. 28, no. 3, pp. 175–185, Jun. 2025, doi: https://doi.org/10.1080/1097198x.2025.2524286.

## Author Biographies

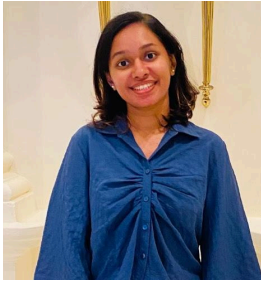

Lakshana Iruni Assalaarachchi is a PhD student at Faculty of Information Technology, Monash University, Australia. Her research interests include software project management, and applications of GenAI in software project management. Assalaarachchi received her Master of Philosophy in Multidisciplinary Studies from University of Sri Jayewardenepura, Sri Lanka. Contact her at lakshana.assalaarachchi@monash.edu.

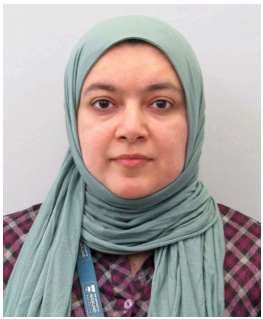

Zainab Masood is an assistant professor at Prince Sultan University, Kingdom of Saudi Arabia. Her research interests focus on the human and socio-technical aspects of software engineering, secure software development (SSD) practices, and the application of artificial intelligence in SSD. Zainab received her PhD in software engineering from University of Auckland, New Zealand. She is a member of the IEEE. Contact her at zmasood@psu.edu.sa.

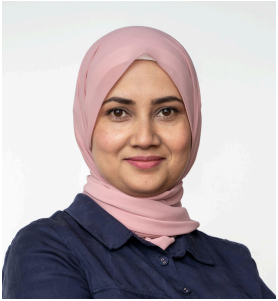

Rashina Hoda is a professor of software engineering at Monash University, Australia. Her research interests include the human and socio-technical aspects of software engineering, artificial intelligence, and digital health. Hoda received her PhD in computer science from Victoria University of Wellington, New Zealand. She is a member of the IEEE. Contact her at rashina.hoda@monash.edu.

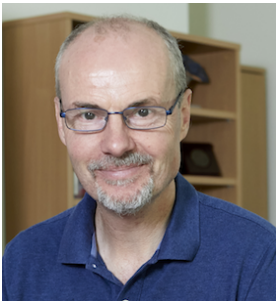

John Grundy is a professor of software engineering at Monash University, Australia. His research interests include human-centric software engineering and automated software engineering. Grundy received his PhD in computer science from University of Auckland, New Zealand. He is a Fellow of the IEEE, Engineers Australia and ASE. Contact him at john.grundy@monash.edu.